\documentclass[useAMS,usenatbib]{mn2e}

\usepackage{graphicx}
\usepackage{times}
\usepackage{natbib}
 
\def\lesssim{\mathrel{\hbox{\rlap{\hbox{\lower4pt\hbox{$\sim$}}}\hbox{$<$}}}}
\def\gtrsim{\mathrel{\hbox{\rlap{\hbox{\lower4pt\hbox{$\sim$}}}\hbox{$>$}}}}

\def\pacman{{\it Pacman }}
\def\be{\begin{equation}}
\def\ee{\end{equation}}
\def\RM{{\rm RM}}

\title[Pacman (I): A new algorithm to calculate RM maps]
{\textbfit{Pacman} (I): A new Algorithm to calculate Faraday Rotation Maps}
\author[K. Dolag, C. Vogt, T. A. En\ss lin]
{K. Dolag$^{1}$\thanks{E-mail: kdolag@pd.astro.it (KD);
cvogt@mpa-garching.mpg.de (CV); ensslin@mpa-garching.mpg.de (TAE)}
C. Vogt$^{2}$ and T. A. En\ss lin$^{2}$ \\ $^{1}$Dipartimento di
Astronomia, Universita di Padova, vicolo dell'Osservatorio 2, 35122
Padova, Italy \\ $^{2}$Max-Planck-Institut f\"{u}r Astrophysik,
Karl-Schwarzschild-Str.1, Postfach 1317, 85741 Garching, Germany}
 
\begin{document}

\date{Accepted ???. Received ???; in original form ???}

\pagerange{\pageref{firstpage}--\pageref{lastpage}} \pubyear{0000}

\maketitle

\label{firstpage}

\begin{abstract} 
We propose a new method to calculate Faraday rotation measure maps
from multi-frequency polarisation angle data. In order to solve the so
called $n\upi$-ambiguity problem which arises from the observationally
ambiguity of the polarisation angle which is only determined up to
additions of $\pm n\upi$, where $n$ is an integer, we suggest using a
global scheme. Instead of solving the $n\upi$-ambiguity for each data
point independently, our algorithm, which we chose to call
\emph{Pacman} (Polarisation Angle Correcting rotation Measure
ANalysis), solves the $n\upi$-ambiguity for a high signal-to-noise
region ``democratically'' and uses this information to assist computations
in adjacent low signal-to-noise areas.
\end{abstract}


\section{Introduction} \label{sec:intro}
Magnetic fields are a common phenomenon on different astrophysical
scales. One possibility to observe astrophysical
magnetic fields and thus, gain insight into their structure, ordering
scale and field strength is given by the Faraday rotation effect. It
arises whenever linearly polarised radio emission passes through a
magnetised medium. The polarisation angle $\varphi$ of the radiation
will be rotated due to the interaction between the magnetic field
component parallel to the propagation direction of radiation and the
radio emission itself. The resulting change in polarisation angle is
proportional to the wavelength squared $\lambda^2$ which can be
expressed by $\varphi = \RM \,\lambda^2 + \varphi^0$, where the
proportionality constant RM in this relation is called Faraday
rotation measure and $\varphi^0$ is the intrinsic polarisation angle
of the emission at the radio source.

The latter two quantities, RM and $\varphi^0$, can be measured through
multi-frequency polarisation observations. The RM can be described as
the line of sight integral over the electron density and the magnetic
field component parallel to the line of sight. Therefore, the RM
contains valuable information on the magnetic field in the foreground
of the polarised radio source. The intrinsic polarisation angle
$\varphi^0$ and thus, the intrinsic magnetic field direction at the
radio source gives insight into the magnetic field configuration at
the source.

The statistical analysis of RM measurements in terms of correlation
functions and equivalently power spectra as developed by
\citet{2003A&A...412..373V} and \cite{2003A&A...401..835E} requires
that the RMs are unambiguously determined. Thus, any ambiguous RM can
lead to misinterpretation of the data investigated. For the
calculation of RM and $\varphi^0$ using the relationship $\varphi =
\RM \,\lambda^2 + \varphi^0$ mentioned above, a least squares fit is
normally applied to the polarisation angle data. Since the measured
polarisation angle $\varphi$ is constrained only to values between 0
and $\upi$ leaving the freedom of additions of $\pm \, n\upi$, where
$n$ is an integer, the determination of RM and $\varphi^0$ is
ambiguous, causing the so called $n\upi$-ambiguity. Therefore, a least
squares fit has to be applied to all possible $n\upi$-combinations of
the polarisation angle data at each data point of the polarised radio
source while searching for the $n\upi$-combination for which $\chi^2$
is minimal.

In principle, $\chi^2$ can be decreased to infinitely small numbers by
increasing RM substantially. \citet{1975A&A....43..233V} and
\citet{1975MNRAS.173..553H} suggested to avoid this problem by
introducing an artificial upper limit for $|\RM|\lid \RM_{\max}$. Since
this is a biased approach, \citet{1979A&A....78....1R} proposed to
assume that no $n\upi$-ambiguity exists between the measurements of
two closely spaced wavelengths taken from a whole wavelength data
set. The standard error of the polarisation measurements is then used
to constrain the possible $n\upi$-combinations for the least squares
fit for the subsequent observed frequencies. We refer to these methods
as the ``standard fit'' algorithms, as they are currently the most
widely used methods. However, these methods might still give multiple
acceptable solutions for data with low signal-to-noise, requiring the
solution to be flagged and all the information carried by these data
is lost. Furthermore, it still can happen that the algorithm chooses a
wrong RM and imprints spurious artefacts on the RM and the $\varphi^0$
maps.

Recently, a completely different approach was proposed by
\citet{2001MNRAS.328..623S} which takes the circular nature of the
polarisation angle into account. The authors apply a maximum
likelihood method to spectral polarisation data. Although this
approach is not biased towards any RM value, it is rather designed for
a large number of observed wavelengths. Similarly, \citet{1996nfra}
and \citet{2004rcfg.procE...6B} propose a RM-synthesis via wide-band
low-frequency polarimetry. However, typically the observations are
only performed at three or four wavelengths especially for extended
(diffuse) radio sources.

Here, we propose a new approach for the unambiguous determination of
RM and $\varphi^0$. We assume that if regions exhibit small
polarisation angle gradients between neighbouring pixels in all
observed frequencies simultaneously, then these pixels can be
considered as connected. Note, that we calculate the gradient modulo
$\upi$, which implies that polarisation angles of 0 and $\upi$ are
regarded as having the same orientation and thus, the cyclic nature of
polarisation angles is reflected. Information about one pixel can be
used for neighbouring ones and especially the solution to the
$n\upi$-ambiguity should be the same. 

In cases of small gradients, assuming continuity in polarisation
angles allows us to assign an absolute polarisation angle for each
pixel with respect to a reference pixel within each observed
frequency. This assignment process has to be done for each spatially
independent patch of polarisation data separately, such as each side
of a double-lobed radio source. The reference pixel is defined to have
a unique absolute polarisation angle and the algorithm will start from
this pixel to assign absolute polarisation angles with respect to the
reference pixel while going from one pixel to its
neighbours. Figuratively, the algorithm eats its way through the set
of available data pixel. It might become clear now why we have chosen
to call the algorithm \emph{Pacman}\footnote{The computer code for
\textit{Pacman} is publicly available at {\tt
http://dipastro.pd.astro.it/\~{}cosmo/Pacman}} (Polarisation Angle
Correcting rotation Measure ANalysis).

\pacman reduces the number of least squares fits in order to solve for
the $n\upi$-ambiguity. Preferably, the reference point is chosen to
have a high signal-to-noise ratio so that in many areas, it is
sufficient to solve the $n\upi$-ambiguity only for a small number of
neighbouring pixels simultaneously and to use this solution for all
spatially connected pixels. Pixels with low signal-to-noise will
profit from their neighbouring pixels allowing a reliable
determination of the RM and $\varphi^0$.

In Sect.~\ref{sec:pacman}, we describe in detail the idea and the
implementation of the \pacman algorithm. In Sect.~\ref{sec:testing},
we test \pacman on artificially generated RM maps and demonstrated its
ability to solve the $n\pi$-ambiguity properly. However, the
application to observational polarisation data and statistical
characterisation of the resulting maps will be presented in the second
paper (Vogt et al.), to which we will refer as Paper II.

\section{The new \textbfit{Pacman} algorithm} \label{sec:pacman}
\subsection{The Idea}

As described in the introduction, the Faraday rotation measure
$\RM_{ij}$ at each point with map pixel coordinate $(ij)$ of the
source, is usually calculated by applying a least squares fit to
measured polarisation angles $\varphi_{ij}(k)$ observed at frequency
$k \in 1...f$ such that
\begin{equation}
\varphi_{ij}(k) = \RM_{ij}\,\lambda_k^2 + \varphi^0_{ij},  
\end{equation}
where $\varphi^0_{ij}$ is the intrinsic polarisation angle at the
polarised source.

Since every measured polarisation angle is observationally constrained
only to a value between 0 and $\upi$, one has to replace
$\varphi_{ij}(k)$ in the equation above by $\tilde{\varphi}_{ij}(k) =
\varphi_{ij}(k) \pm n_{ij}(k)\,\upi$, where $n_{ij}(k)$ is an integer,
leading to the so called $n\upi$-ambiguity. Taking this into account,
a least squares fit to calculate RM$_{ij}$ and $\varphi^0_{ij}$ at
each pixel has to be applied by allowing all possible combinations of
$n_{ij}(k)\upi$ while determining the $n_{ij}(k)$ for which the
$\chi_{ij}^2$ is minimal. The presence of observational noise might
cause a standard least squares fit, as suggested by
\citet{1975A&A....43..233V}, \citet{1975MNRAS.173..553H} or
\citet{1979A&A....78....1R}, to choose a spurious RM value especially
for areas of low signal-to-noise.

The idea of \pacman is to reduce the number of pixels for which the
$n\upi$-ambiguity has to be individually solved. This is done by
splitting the solution of the $n\upi$-problem into two problems, a
local and a global one,
\begin{equation}
n_{ij}(k) = \tilde{n}_{ij}(k) + n(k),   
\end{equation}
where $\tilde{n}_{ij}(k)$ is the local solution, linking polarisation
angles of neighbouring pixels within a frequency map, and $n(k)$ is
the global solution to the problem, linking polarisation angles of the
different frequencies. The local part $\tilde{n}_{ij}(k)$ is
determined by construction of absolute polarisation angle maps for
each frequency with respect to a high signal-to-noise reference pixel
being defined to possess a unique polarisation angle. The term
absolute polarisation angle is to be understood as a value determined
relative to the reference pixel by adding $\pm n\upi$ to the measured
polarisation angles in order to remove jumps of the order of $\upi$ in
the measured polarisation angle map of each observed frequency. The
global $n\upi$ ambiguity is solved for a high signal-to-noise area
surrounding the reference pixel resulting in $n(k)$. This is then also
the solution of the global $n\upi$-problem for all spatially connected
points which are assigned absolute polarisation angles with respect to
this reference area.

The splitting of the problem in a local and a global one is possible
if the real polarisation angle $\bar{\varphi}_{ij}(k)$
\mbox{($\bar{\varphi}_{ij}(k) = \RM_{ij}\,\lambda_k ^2 +
\varphi^0_{ij}$)} is a smooth quantity which does not change more than
$\pm \upi/2$ between neighbouring pixels. In
Sect.~\ref{sec:pixreject}, it is described how \pacman ensures that
only pixels fulfilling this condition are used.

The source might consists of several spatially independent areas of
polarisation, to which we will refer as patches in the following. The
$n\upi$-ambiguity has to be solved for each of these patches
separately, requiring separate reference pixels to be defined for the
construction of absolute polarisation angles.

The advantage of \pacman is that the global $n\upi$-ambiguity is
solved only for pixels having the highest signal-to-noise
ratios. Therefore, noisier pixels which are situated at the margin of
the source profit from an already defined global solution $n(k)$ to
the $n\upi$-ambiguity, making a reliable determination of RM$_{ij}$
and the intrinsic polarisation angle $\varphi^0_{ij}$ for these pixels
possible.

\subsection{The Basic Algorithm}\label{sec:pixreject}
\begin{figure}
\includegraphics[width=0.5\textwidth]{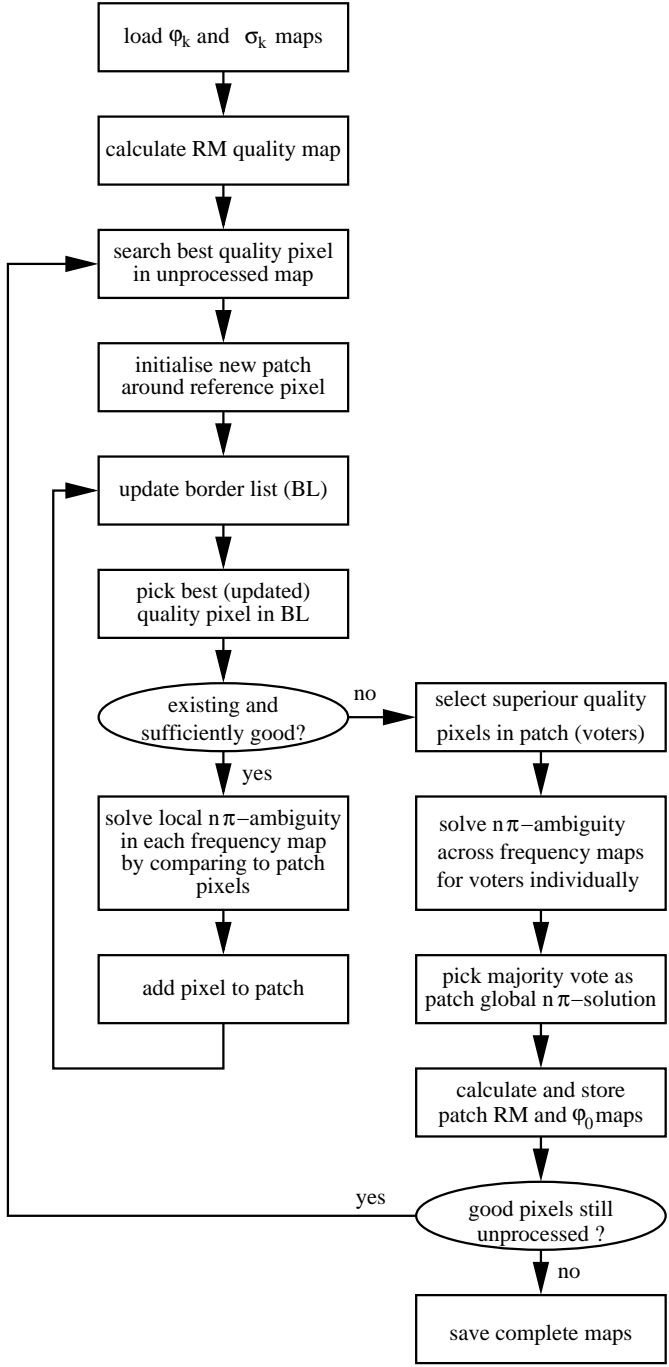}
\caption{A flow chart representing the individual steps involved in
the calculation of RM and $\varphi^0$ maps as performed by our
algorithm \textit{Pacman}.}
\label{fig:flowchart}
\end{figure}

The algorithm starts with reading the maps of polarisation angles and
of its errors, processes them, and when finished, it saves the
different calculated maps. A flow chart of our \pacman algorithm is
exhibited in Fig.~\ref{fig:flowchart}, which shows schematically the
procedure the algorithm follows in order to determine the solution to
the $n\upi$-ambiguity and to calculate the various maps.

After loading of the various polarisation data for the different
frequencies, the algorithm calculates a quality measure map. Different
actions are involved in the processing of the data. One of the
processes is building a patch which involves the construction of
absolute polarisation angle maps. First, the reference pixel for a
patch of polarisation is determined from the quality measure map. This
step is followed by solving the local $n\upi$-ambiguity in each
frequency map for pixels being spatially connected to the defined
patch reference pixel and having small gradients in polarisation angle
to the neighbouring pixels at all frequencies. This process is mainly
governed and stopped by quality requirements.

When the patch building process is stopped then the algorithm solves
the global $n\upi$-ambiguity across the frequency maps individually
for a selected set of superior quality pixels (voters) within the
patch. The majority of votes is used as the global patch
$n\upi$-solution and the resulting RM and $\varphi^0$ values are
stored. If there are still unprocessed high quality pixel left, the
algorithm starts again to build a new patch. Otherwise \pacman
finishes and saves the resulting complete maps.

The values for RM and $\varphi^0$ are calculated following a weighted
least squares fit expressed by
\begin{eqnarray}
\RM_{ij} & = &
\frac{\overline{S_{ij}}\,\overline{\lambda^2\varphi_{ij}} -
\overline{\lambda^2}\,\overline{\varphi_{ij}}}
{\overline{S_{ij}}\,\overline{\lambda^4} - {\overline{\lambda^2}}^2} \\
\varphi^0_{ij} & = &
\frac{\overline{\varphi_{ij}}\,\overline{\lambda^4} -
\overline{\lambda^2}\,\overline{\lambda^2\varphi_{ij}}}
{\overline{S_{ij}}\,\overline{\lambda^4} - {\overline{\lambda^2}}^2}
\end{eqnarray}
where \mbox{$\overline{S_{ij}} = \sum _{k=1} ^f 1/\sigma^2_{k_{ij}}$},
\mbox{$\overline{\varphi_{ij}} = \sum _{k=1} ^f
\varphi_{ij}(k)/\sigma^2_{k_{ij}}$}, \mbox{$\overline{\lambda^2} =
\sum _{k=1} ^f \lambda_k ^2/\sigma^2_{k_{ij}}$},
\mbox{$\overline{\lambda^2 \varphi_{ij}} = \sum _{k=1} ^f
\lambda_k^2\,\varphi_{ij}(k)/\sigma^2_{k_{ij}}$} and
\mbox{$\overline{\lambda^4} = \sum _{k=1} ^f \lambda_k
^4/\sigma^2_{k_{ij}}$}. In these relations, $\sigma_{k_{ij}}$ is the
standard error of the polarisation angle $\varphi_{ij}(k)$ at the
pixel coordinate $(ij)$ at the $k$th wavelength $\lambda_k$.

Since the performance of the algorithm is mainly governed by quality
requirements, we need some quality measure in order to rank the map
pixels. One good candidate is the expected uncertainty
$\sigma^{\RM}_{ij}$ of any RM value obtained in an error weighted
least squares fit which is calculated by
\begin{equation}
\label{eq:deltaRM}
\sigma_{ij}^{\RM} =
\sqrt{\frac{\overline{S_{ij}}}
{\overline{S_{ij}}\, \overline{\lambda^4} - \overline{\lambda^2}^2}},
\end{equation} 
where the terms are defined as above.

The uncertainty $\sigma^{\RM}_{ij}$ of the RM value is used to assign
each pixel an initial quality $q_{ij} = \sigma^{\RM}_{ij}$ since
$\sigma^{\RM}_{ij}$ accounts for the statistic used to determine the
RM maps. Hence, small values of $q_{ij}$ indicate high quality
pixels. The quality is then used to determine the way \pacman goes
through the data which will be preferably from high to low quality
data pixels.

For the construction of absolute polarisation angle maps for each
frequency, \pacman starts at the best quality pixel having the
smallest value $q_{ij}$, which is defined to be the reference
pixel. For this first point, the measured polarisation angle
$\varphi_{ij}(k)$ is defined to possess a unique polarisation angle
value $\bar{\varphi}_{ij}(k)=\varphi_{ij}(k)$ for each observed
frequency. It is important to note that the reference pixel is the
same for all frequencies.

Then, \pacman compiles what we call the border list (BL). It contains
pixels being direct neighbours to points which have been already
assigned an absolute polarisation angle. In the following, the set of
adjacent pixels ($i \pm 1\,j \pm 1$) to the pixel ($i j$) will be
referred to as direct neighbours ($i'j'$). Beginning with the direct
neighbours of the reference pixels the border list is continuously
updated during the progression of the algorithm. A pixel can be
rejected from the border list if the standard error $\sigma_{k_{ij}}$
of the polarisation angle for any frequency at this pixel exceeds a
certain limit $\sigma_k ^{\max}$ which is set at the beginning of the
calculation. However, this requirement can be relaxed (see
Sect.~\ref{sec:multifrequency}).

Having defined the reference pixel, \pacman assigns absolute
polarisation angles to pixels within the border list always starting
with the pixel having the best quality, i.e. the lowest $q_{ij}$
value. For this pixel, the algorithm solves the local
$n\upi$-ambiguity with respect to $\tilde{n}_{ij}(k) \in \bmath{Z} $
by minimising the expression
\begin{equation}
\label{eq:localamb}
\sigma\, ^{\Delta} _{ij} = \sum _{\{i'j'\}\,\in\,D_{ij}} \left[ \left(
\varphi_{ij}(k) \pm \tilde{n}_{ij}(k) \upi \right) -
\bar{\varphi}_{i'j'}(k) \right] ^2,
\end{equation}
where $D_{ij}$ is the set of all direct neighbours $(i'j')$ to the
pixel $(ij)$ which have already been assigned an absolute polarisation
angle. The resulting $\tilde{n}_{ij}(k)$ value determines the absolute
polarisation angle $\bar{\varphi}_{ij}(k) = \varphi_{ij}(k)
\pm \tilde{n}_{ij}(k) \upi$ which has the smallest difference to the
already defined absolute polarisation angles of adjacent pixels
$\bar{\varphi}_{i'j'}(k)$. For each pixel, this is done at each
frequency $k \in 1....f$ independently but simultaneously. Thus, \pacman
goes the same way through the data in each frequency. Note that one
can introduce a value $\sigma^{\Delta}_{\max}$ which causes \pacman to
reject the pixel if $\sigma\, ^{\Delta}_{ij}$ exceeds
$\sigma^{\Delta}_{\max}$ (see
Sect.~\ref{sec:spurpoints}). Additionally the border list is updated
to include direct neighbours to the recently processed pixel $(ij)$
which have not yet been assigned an absolute polarisation angle.
 
\pacman repeats this process for the best remaining pixel in the border
list which has the lowest $q_{ij}$ and so on, until the whole patch
consists of spatially connected pixels with assigned absolute
polarisation angles and no acceptable neighbouring pixels remain in
the border list.

At this stage, \pacman solves the global $n\upi$-ambiguity $n(k)$ by
applying a standard least squares fit to a set of best constrained
pixels within the patch (i.e. the surrounding area of the reference
pixel). \pacman solves for each of these best constrained pixels the
global $n\upi$-ambiguity independently by either using the method of
\citet{1975MNRAS.173..553H, 1975A&A....43..233V} or of
\citet{1979A&A....78....1R} minimising the $\chi^2$ for all possible
combinations of $n\upi$. It searches then for the solution $n(k)$,
which has been derived for the majority of pixels. This democratically
determined solution $n(k)$ is used to perform a least squares fit of
the other pixels defined within this patch. Note that in an extreme
case, only the reference pixel could be used to solve the global
$n\upi$-ambiguity.

If there are still good quality pixels left in the map which were not
connected to the first patch of good quality data, the algorithm
begins a new patch of absolute polarisation angles, starting again
with the best remaining pixel with smallest $q_{ij}$. A new patch of
absolute polarisation angles is constructed. As a result, the source
will be divided in a set of spatially independent polarisation
patches, and the global $n\upi$-ambiguity is solved once for each
patch separately. A $q_{ij}^{\rm {thresh}}$ can be introduced to
prevent \pacman from starting new patches if the pixels remaining have
quality values above that threshold value whose value is chosen in the
beginning.

\subsection{Improving Quality}
The solution of the local $n\upi$-ambiguity for pixel $(ij)$ becomes
more reliable as more direct neighbours of these pixels have their
absolute polarisation angle $\bar{\varphi}_{i'j'}$ defined. This is
taken into account by modifying the quality $q_{ij}$ of the pixels
added to the border list to
\begin{equation} 
\label{eq:modqual}
1/q_{ij} = 1/\sigma^{RM} _{ij} +
\frac{\alpha}{n_{D_{ij}}^{\beta}}\left(\sum _{i'j'
\,\in\,D_{ij}}{1/\sigma^{RM} _{i'j'}} \right),
\end{equation}
where $n_{D_{ij}}$ is the number of already defined neighbours
$(i'j')$, and $\alpha$ and $\beta$ are free parameters. This ensures
that pixels having more already defined neighbours are considered
before others although the original data point might have a smaller
signal quality, i.e a higher $q_{ij}$. Values between 0 and 1 for the
free parameter $\beta$ are reasonable and yield good results. We used
$\alpha = \beta = 1$ and observed that \pacman goes through the
acceptable data points of the polarisation patches in a uniform
manner.

\subsection{Restricting Gradients}
The algorithm might be faced with a situation where a high quality
pixel can be influenced by a very poor quality pixel, for example when
the polarised radio source consists of two radio lobes each having a
good signal-to-noise ratio which are connected by a bridge containing
only low quality pixels. The \pacman algorithm would start by defining
absolute polarisation angles from one of the two lobes, eventually
reaching the bridge of low signal-to-noise and entering the second
lobe from there. In such cases, it might happen that within the area
of low quality data pixels a distinct determination of absolute
polarisation angles by solving the local $n\upi$-ambiguity using
Eq. (\ref{eq:deltaRM}) is no longer possible. The entire second lobe
would then suffer, and wrong solutions would be introduced.

In order to avoid such situations, the algorithm is restricted to
accept in the border list only neighbours which have a lower quality
than the one under consideration. This forces the algorithm to go
always from high to low quality pixels, leading to an artificial
splitting of connected regions in the map into different patches.

However, such a strict rule would lead to heavy fragmentation and is
not preferable. Therefore, we introduced a parameter $g$ to relax this
strict limitation such that a new pixel is only accepted in the border
list when the relation
\begin{equation}  
q_{i'j'} > g * q_{ij}  
\end{equation}
is fulfilled. This relation is always accounted for when
adding a new pixel to the border list. We found that $g$ between 1.1
and 1.5 is a good choice.

\subsection{Topological Defects}
The algorithm can also be faced with situations which we call
topological defects. These defects can be understood by supposing a
ring-like polarised structure for which all polarisation angles point
towards the centre of the ring. Starting at any point of the ring to
define absolute polarisation angles and following the ring structure,
a jump will appear on the border of the first to the last defined
absolute polarisation angles after having performed a full
circle. \citet{1977MNRAS.180..163S} describe this problem
which they encountered by their analysis of polarisation data.

In observational data, topological defects are often more complex
structured. The artificial jump introduced will cause difficulties in
the solution of the local $n\upi$-ambiguity because this procedure
relies on already defined absolute polarisation angles of neighbouring
pixels. When confronted with this situation, our algorithm divides the
list of direct neighbours into sublists possessing similar
polarisation angles. The sublist containing the most pixels is then
used to assign the absolute polarisation angle by solving the local
$n\upi$-ambiguity for the pixel under consideration.

The locations of the polarisation angle steps of topological defects
are somewhat artificial since they depend on the actual path of the
algorithm through the data. Since the algorithm processes all
frequencies simultaneously, the steps of topological defects are at
the same positions within all frequency maps, and therefore do not
cause any further problems. However, if \textit{Pacman} encounters a
jump in the polarisation angle to all possible neighbour sublists in
any of the frequencies under consideration, \pacman will reject this
pixel and this pixel will not be considered for this patch but queued
back for consideration for the next patches to be constructed. In our
experience, these topological defects are rare events, but it is
necessary to take them into account for the solution of the local
$n\upi$-ambiguity.

\subsection{Spurious Points\label{sec:spurpoints}}
The intrinsic polarisation angle distribution might show strong
gradients extending over a few pixels.  This could lead to situations,
where our algorithm can not solve the local $n\upi$-ambiguity at all
frequencies simultaneously. Therefore, \pacman refuses to assign an
absolute value $\bar{\varphi}_{ij}$ to pixels when Eq.~(4) yields a
value $\sigma^{\Delta} _{ij}$ above a certain threshold
$\sigma^{\Delta} _{\max}$, which can be set at the beginning of the
calculation.

The pixels in the regions where this might occur most often have a low
signal-to-noise ratio. In our experience, such situations always occur
at strongly depolarised areas, leading to blanked regions in the RM
distribution.

\subsection{Multi-Frequency Fits\label{sec:multifrequency}}
The aim of any RM derivation algorithm should be to calculate RMs for
an area as large as possible using as much information as is
available. On the other hand, for radio observations the total radio
intensity decreases with increasing frequency. This can lead to the
problem that the area of acceptable polarisation data at a high
frequency is much smaller than at a lower frequency. This is
especially true for (diffuse) extended radio sources. Furthermore, the
limit of allowed standard errors $\sigma_{k} ^{\max}$ of the
polarisation angle might be exceeded for only one frequency leaving the
values for the other frequencies still in the acceptable
range. 

\pacman can take this into account and performs the RM fit by omitting
the polarisation angles at frequencies which do not meet the quality
requirements. In order to do that, \pacman uses the standard errors
$\sigma_{k_{ij}}$ of polarisation angles to define independently for
each frequency, areas in which the RM map is permitted to be
produced. An additional parameter $k_{\rm min} \lid f$ is introduced
which describes the minimum number of frequencies allowing the freedom
to use any combination of the minimum of frequencies. 

In cases of $k_{\rm {min}} < f$, \pacman will start to determine the
solution to global and local $n\upi$-ambiguity only for the pixels
fulfilling the quality requirements at all $f$ frequencies. After
finishing that, the algorithm proceeds to include pixels satisfying
the quality criteria in less than $f$ frequencies. For these pixels,
the same patch building procedure as described in
Sect.~\ref{sec:pixreject} applies with some modification in order to
prevent the final maps from heavy fragmentation. The best quality
pixel among the remaining pixel is picked but before starting a new
patch, \pacman tests if the pixel under consideration adjoins a patch
which has already been processed before. If the pixel adjoins such a
patch, \pacman tries to solve the local $n\upi$-ambiguity following
Eq.(\ref{eq:localamb}) and applies the patch solution of the global
$n\upi$-ambiguity to the pixel. If the pixel is neither adjoined to an
already processed patch nor the local $n\upi$-ambiguity solvable
(i.e. $\sigma_{ij}^{\Delta}$ exceeds $\sigma_{\rm {max}}^{\Delta}$)
then a new patch is initialised by this pixel.

Another possibility is to force \pacman to use certain frequencies in
all circumstances. Thus, if the quality requirements are not fulfilled
for these particular frequencies, there will be no RM value determined
for the pixel under consideration.  This has the advantage that one
can use relatively close frequencies as a basis and then include other
frequencies at points when a reliable polarisation signal is
detected. The advantage of this technique is discussed in Paper II.

\subsection{Additional Information}
Apart from the resulting RM and $\varphi^0$ maps, \pacman provides
sets of additional information about the data in order to estimate the
reliability of the results. A patch map which contains all patches
used is one example. Such a map is very useful, especially if one
requires a minimal number of pixels in a patch in order to accept any
calculated RM values from a particular patch. A map which includes
rejected and thus flagged pixels, can also be obtained. Probably more
important are the final $\chi^2$- and $\sigma^{\RM}$- maps which are
also provided by \emph{Pacman}. This information allows one to
understand the reliability of the RM maps obtained and can be used for
further evaluation and analysis of the RM maps.

\begin{figure*}
\resizebox{\hsize}{!}{\includegraphics{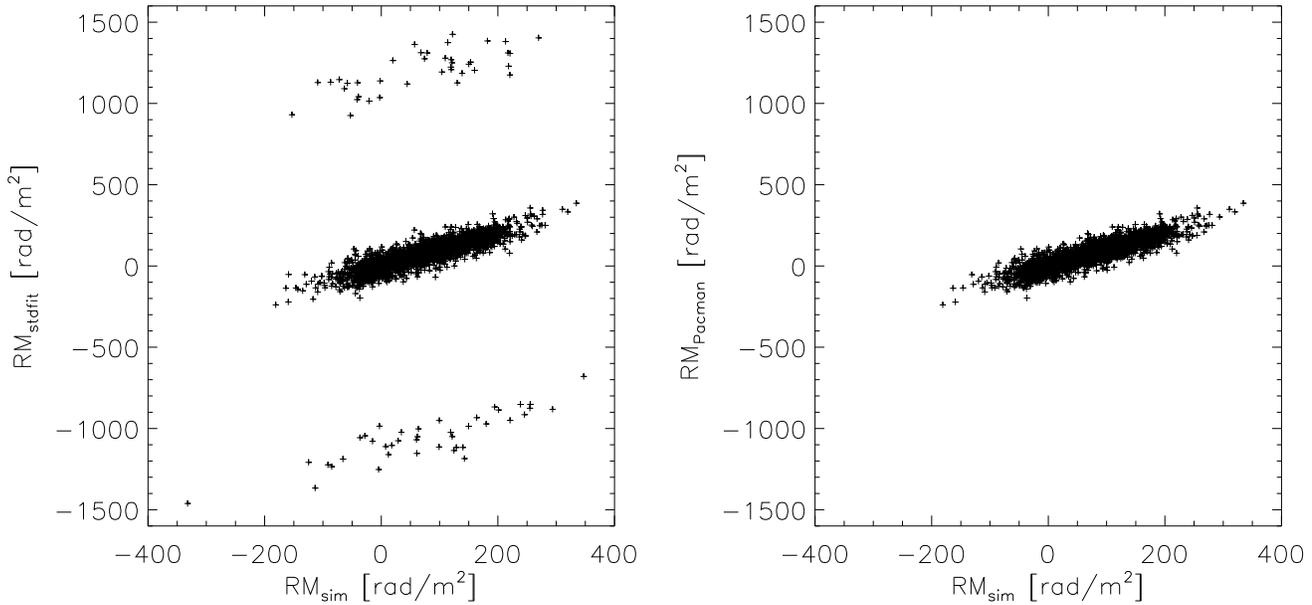}}
\caption[]{Comparison between the simulated $\RM_{{\rm sim}}$
values for an artificially generated set of polarisation angle maps of
four frequencies and values derived by \pacman $\RM_{{\rm pacman}}$
and the standard fit algorithm $\RM_{{\rm stdfit}}$. The scatter is
due to noise which was added to the polarisation angle maps. Note the
wrong solutions of the $n\pi$-ambiguity results which were calculated
by the standard fit indicated by the points around $\pm$ 1000 rad
m$^{-2}$.}
\label{fig:compare}
\end{figure*}

\section{testing the \textbf{\textit{Pacman}}
algorithm}\label{sec:testing} 
In order to demonstrate the ability of Pacman to solve the
$n\pi$-ambiguity, we use artificially generated maps. To generate
them, we start from a polarisation data set from Abell 2255
\citep{2002taiwan.conf}, which was kindly provided by Federica
Govoni. We calculated the $\RM$ and $\varphi_0$ maps from this data
set and assume that these two maps consist of exact values. We
generated then polarisation maps at four different frequencies which
would exactly result from the $\RM$ and $\varphi_0$ maps. As error
maps, we generated Gaussian deviates for each frequency which were
multiplied by the original error maps provided by Federica Govoni. The
so generated error maps were added to the generated frequency maps in
order to provide realistic mock observations.

\pacman and the standard fit algorithm were used to calculate
the corresponding 'observed' $\RM$ and $\varphi_0$ maps which then
were compared pixel by pixel to the initial exact maps. The result of
this comparison is shown in Fig.~\ref{fig:compare}. On the right
panel, the comparison between the $\RM_{{\rm pacman}}$ values of the
\pacman map and the values $\RM_{{\rm sim}}$ of the initial map is
shown. The scatter in the data is due to the noise which was added to
the frequency maps. On the left panel of Fig.~\ref{fig:compare}, the
pixel by pixel comparison between the values of the standard fit
$\RM_{{\rm stdfit}}$ map and the initial $\RM_{{\rm sim}}$ map is
shown. Again the scatter is due to the added noise. However, one can
clearly see the points at $\pm$ 1000 rad m$^{-2}$ which deviate from
the initial data and are due to the wrongly solved $n\pi$-ambiguity in
the case of the standard fit. Thus, this test demonstrates that
\pacman yields reliable results to the $n\pi$-ambiguity.

\section{Conclusions} \label{sec:conclusion}
We have presented a new algorithm for the calculation of Faraday
rotation maps from multi-frequency polarisation data sets. Unlike
other methods, our algorithm uses global information and connects
information about individual neighbouring pixels with one another. It
assumes that if regions exhibit small polarisation angle gradients
between neighbouring pixels in all observed frequencies
simultaneously, then these pixels can be considered as connected, and
information about one pixel can be used for neighbouring ones, and
especially, the solution to the $n\upi$-ambiguity should be the
same. We like to stress, that this is a very weak assumption, and --
like all other criteria used within the \pacman algorithm -- only
depends on the polarisation data at hand and the signal-to-noise ratio
of the observations.

Our \pacman algorithm is especially useful for the calculation of RM
and $\varphi^0$ maps of extended radio sources. Global algorithms as
implemented in \pacman are preferable for the calculation of RM maps
and needed if reliable RM values are desired from low signal-to-noise
regions of the source.

Our \pacman algorithm reduces $n\upi$-artefacts in noisy regions and
makes the unambiguous determination of RM and $\varphi^0$ in these
regions possible. \pacman allows to use all available information
obtained by the observation. Any statistical analysis of the RM maps
will profit from these improvements.

A more detailed description of the application of \pacman to
observations and various statistical tests performed on resulting RM
maps in comparison to the standard fit algorithms are given in Paper
II.

\section*{acknowledgements}
We want to thank Federica Govoni for allowing us to use their data on
Abell 2255 to test the new algorithm. We like to thank Tracy Clarke,
James Anderson, Phil Kronberg and an anonymous referee for useful
comments on the manuscript. K.~D.~acknowledges support by a Marie
Curie Fellowship of the European Community program "Human Potential"
under contract number MCFI-2001-01227.

\bibliographystyle{mn2e}
\bibliography{literature}

\label{lastpage}

\end{document}